\documentclass[superscriptaddress,twocolumn,prb,longbibliography]{revtex4-1}
\usepackage{graphicx}			
\usepackage{dcolumn}			

\usepackage{verbatim}
\usepackage{color}

\usepackage{bm}					
\usepackage[mathlines]{lineno}	
\usepackage{amsmath}

\usepackage{natbib}
\newcommand{\und}[1]{_{\mathrm{#1}}}

\newcommand{\Ah}{\mathcal{A}_{\mathrm{high}}}
\newcommand{\Al}{\mathcal{A}_{\mathrm{low}}}
\newcommand{\dm}{\delta_{\rm m}}

\begin{document}

\title{{\bf Optically levitated nanoparticle as a model system for stochastic bistable dynamics}}%

\author{F. Ricci}%
\affiliation{ICFO-Institut de Ciencies Fotoniques, The Barcelona Institute of Science and Technology, 08860 Castelldefels (Barcelona), Spain}

\author{R. A. Rica}%
\affiliation{ICFO-Institut de Ciencies Fotoniques, The Barcelona Institute of Science and Technology, 08860 Castelldefels (Barcelona), Spain}

\author{M. Spasenovi\'c}%
\affiliation{ICFO-Institut de Ciencies Fotoniques, The Barcelona Institute of Science and Technology, 08860 Castelldefels (Barcelona), Spain}

\author{J. Gieseler}%
\affiliation{ICFO-Institut de Ciencies Fotoniques, The Barcelona Institute of Science and Technology, 08860 Castelldefels (Barcelona), Spain}
\affiliation{Physics Department, Harvard University, Cambridge, Massachusetts 02138, US.}

\author{L. Rondin}%
\affiliation{ETH Z\"urich, Photonics Laboratory, 8093 Z\"urich, Switzerland}

\author{L. Novotny}%
\affiliation{ETH Z\"urich, Photonics Laboratory, 8093 Z\"urich, Switzerland}

\author{R. Quidant}%
\affiliation{ICFO-Institut de Ciencies Fotoniques, The Barcelona Institute of Science and Technology, 08860 Castelldefels (Barcelona), Spain}
\affiliation{ICREA-Instituci\'o Catalana de Recerca i Estudis Avan¸cats, 08010 Barcelona, Spain}

\maketitle

{\bf  Nano-mechanical resonators have gained an increasing importance in nanotechnology owing to their contributions to both fundamental and applied science. Yet, their small dimensions and mass raises some challenges as their dynamics gets dominated by nonlinearities that degrade their performance, for instance in sensing applications. Here, we report on the precise control of the nonlinear and stochastic bistable dynamics of a levitated nanoparticle in high vacuum. We demonstrate how it can lead to efficient signal amplification schemes, including stochastic resonance. This work paves the way for the use of levitated nanoparticles as a model system for stochastic bistable dynamics, with applications to a wide variety of fields.}
\newline
\newline
\noindent{\bf Introduction}
\newline Nano-mechanical resonators need to meet criteria of light mass and high-Q factor in order to maximize their performances when operated as linear force transducers. However, these features lead to intrinsically nonlinear behaviour~\cite{Gieseler2013Thermal}, with consequent vanishing dynamical range. To overcome this limitation, modern nanotechnology requires new sensing schemes that take nonlinearities into account and even benefit from them~\cite{Villanueva2013Surpassing}. Many of the proposed solutions operate inside an instability region~\cite{Aldana2014Detection,Papariello2016Ultrasensitive} or close to a bifurcation point~\cite{Karabalin2011Signal,Siddiqi2004RF-Driven}, where the system ideally becomes infinitely sensitive. Others exploit fluctuations of noisy environments to trigger stochastic resonances~\cite{Gammaitoni1998Stochastic} that amplify weak harmonic signals~\cite{Venstra2013Stochastic,Badzey2005Coherent,Almog2007Signal}. For all these sensing applications, a single nanopartticle optimally decoupled from the environment represents a particularly interesting system in the family of high-Q resonators. In fact, since its very first realization~\cite{Gieseler2012Subkelvin}, optical levitation of a nano-scale object in vacuum has enabled several ground-breaking experiments, including demonstration of zeptonewton force sensitivity~\cite{Gieseler2013Thermal,Ranjit2016Zeptonewton}, tests of fluctuation theorems and stochastic thermodynamics~\cite{Gieseler2014Dynamic,Millen2014Nanosacale}, as well as the observation of photon recoil heating~\cite{Jain2016Direct}.
\newline \indent In this work, we demonstrate full control on the linear, non-linear and bistable dynamics of a levitated nanoparticle in high vacuum and under the effect of external noise. The potential of our platform is validated by the implementation of two nonlinear amplification schemes, including stochastic resonance~\cite{Gammaitoni1998Stochastic}. Remarkably, we demonstrate, in excellent agreement with theory, up to $\sim 50~\rm dB$ amplification of non-resonant harmonic force at the atto-newton scale. In addition, the unprecedented level of control achieved will enable the use of levitated nanoparticles as a model system for stochastic bistable dynamics with applications to a wide range of fields including biophysics~\cite{Hayashi2012Single,Angeli2004Detection}, chemistry~\cite{Simakov2013Noise,Hanggi1990Reaction-rate} and nanotechnology~\cite{Myers2015Information}.

\vspace{.5cm}
\noindent{\bf Results}
\begin{figure*}[ht]
\centering
\includegraphics{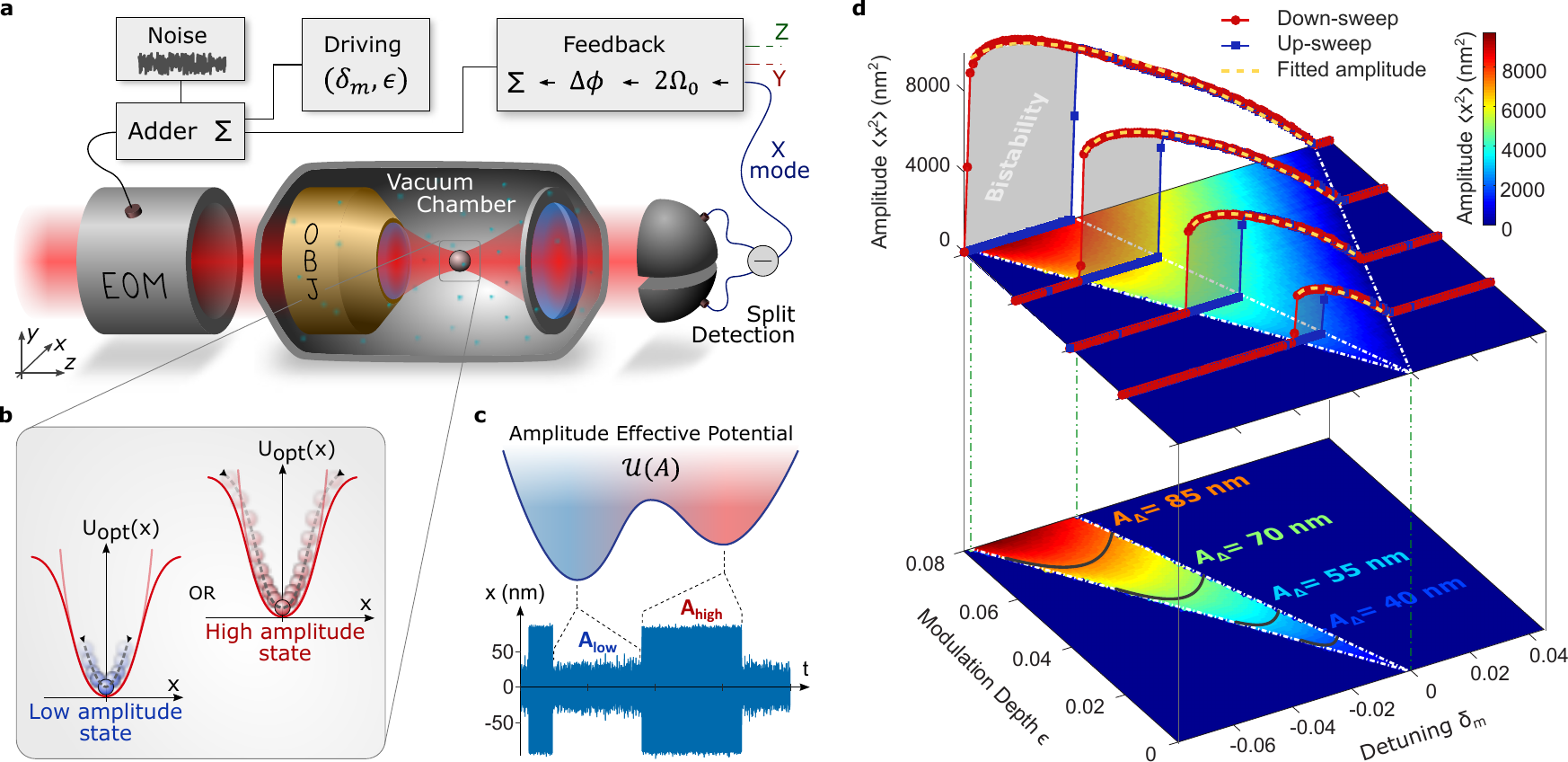}
\caption{\footnotesize \textbf{Experimental set-up and nonlinear response.} (\textbf{a}) A microscope objective (OBJ) focuses a laser beam inside a vacuum chamber, where a single silica nanoparticle is trapped. Its motion is measured with a split detection scheme and parametric feedback is applied via an electro-optical modulator (EOM) in order to cool its center of mass motion. (\textbf{b}) A linear low amplitude and a nonlinear high amplitude oscillation states coexist when bistability is induced by the coherent driving. (\textbf{c}) The bistable dynamics can be modelled by an effective amplitude double-well potential. Injected optical noise activates stochastic switching between the two states. (\textbf{d}) (top) Measured 2D false color map representing the particle amplitude response in the driving parameter phase space $(\delta_m,\epsilon)$. The red and blue cross-cuts, correspond respectively to down- and up- sweeps, and emphasize the typical hysteretic non-linear response of the resonator. Yellow lines are fits to the high amplitude solution $A(\dm)$ and allow to retrieve the nonlinear coefficients $\eta$ and $\xi$. The bistable regime (bottom map), can be isolated subtracting up-sweep from down-sweep maps. Solid lines correspond to iso-amplitude lines~(\ref{eq:iso-amplitude}): subsets of the parameter space where $\Ah$, and consequently $\mathcal{A}_{\Delta}=\Ah-\Al$, are kept constant.}
\label{fig:Exp_SetUp}
\end{figure*}
\newline \indent Our experiment is sketched in Fig.~\ref{fig:Exp_SetUp}a. A single silica nanoparticle ($d\sim177~\rm{nm}$ in diameter) is optically trapped in vacuum by a tightly focused laser beam, and its 3D trajectory monitored with a balanced split detection scheme. The detector signal controls an electro-optic modulator (EOM) that feedback-cools all three translational degrees of freedom by modulating the laser intensity. In high-vacuum conditions ($P\sim 10^{-6}~\rm{mbar}$) the mechanical Q-factor, given by the gas damping $\gamma_0$, is measured to be $Q=1.2 \times 10^{8}$. Under feedback, the nanoparticle oscillates at non-degenerate angular frequencies $\Omega_{0}^{(x,y,z)}$ with very small amplitudes, corresponding to sub-Kelvin effective temperatures~\cite{Gieseler2012Subkelvin}. By optimizing the feedback settings (Supplementary Note~2) and carefully screening important sources of noise such as mechanical vibrations and air turbulences, we obtain a highly stable resonator with frequency fluctuations improved by one to two orders of magnitude compared to previous works~\cite{Gieseler2013Thermal,Gieseler2014Nonlinear} (See Supplementary Note~3). When one of the spatial modes is parametrically driven at resonance~\cite{Gieseler2014Nonlinear}, the particle explores the anharmonic part of the optical potential, which can be modelled as a Duffing nonlinearity~\cite{Gieseler2013Thermal}. As a result, the equation of motion for the driven coordinate, which we choose to be $x$, reads
\begin{equation} \label{eq:motion_02}
\ddot x  + \left( \gamma_0 + \Omega_0 \eta x^2 \right)\dot{x} + \Omega_0^2 \left[1 + \xi x^2 + \epsilon \cos \left( \Omega_m t \right)\right]x = \frac{\mathcal{F}}{m}
\end{equation}
where $\eta$ is the non-linear feedback-induced damping coefficient, $\xi$ is the Duffing term prefactor, $\epsilon$ and $\Omega_{m}$ are, respectively, the modulation depth and modulation frequency of the driving signal. Finally, the fluctuating force $\mathcal{F}=\mathcal{F}\und{th}+\mathcal{F}\und{noise}$ has two contributions. The first one, $\mathcal{F}\und{th}$, represents the stochastic force arising from random collisions with residual air molecules in the chamber, while the second, $\mathcal{F}\und{noise}$, represents artificial parametric noise that we add through the EOM. See Supplementary Note~1 for further experimental set-up details. 
\newline \indent The nonlinear response of the resonator is fully characterized by measuring the oscillation amplitude in the driving parameters phase space $(\dm,\epsilon)$, where $\delta\und{m}=\Omega\und{m}/\Omega_0-2$ is the normalized detuning~\cite{Gieseler2014Nonlinear}. When the modulation frequency is decreased (down-sweep) across the resonant condition $\delta_m = 0$, the amplitude response spans a triangular region corresponding to the first instability tongue~\cite{Lifshitz2008Nonlinear}. This is shown in the top 2D false color map of Fig.~\ref{fig:Exp_SetUp}d. In the opposite case of a frequency up-sweep, however, the system displays hysteresis, and the tongue results in a narrower width. The red and blue cross cuts in Fig.~\ref{fig:Exp_SetUp}d clearly show this behaviour, identifying a region where two stable oscillation states coexist (see Fig.~\ref{fig:Exp_SetUp}b,c). The bistability region, shown at the bottom of Fig.~\ref{fig:Exp_SetUp}d, is obtained by subtracting up- from down-sweeps maps. By fitting our experimental data with the analytical solution $A(\delta\und{m},\epsilon)$ (See Supplementary Equation~1 and Ref.~\cite{Gieseler2014Nonlinear}) we retrieve the nonlinear coefficients $\eta=(18.1\pm0.4)~\mathrm{\mu m}^{-2}$ and $\xi=(-9.68\pm0.15)~\mathrm{\mu m}^{-2}$. Such precision measurements are made possible by the highly stable oscillation frequencies that lead to minimal drifts of the detuning parameter ($\Delta\delta_{\rm m} \approx 2\Delta\Omega / \Omega_0 \lesssim 10^{-3}$, over the over the entire measurement time), therefore allowing to predict with excellent agreement the boundaries of the instability tongues (white lines in Fig.~\ref{fig:Exp_SetUp}d). Further details on spectral features of the nonlinear response are provided in Supplementary Note~4.
\begin{figure*}[ht]
\centering
\includegraphics[width=0.85\textwidth]{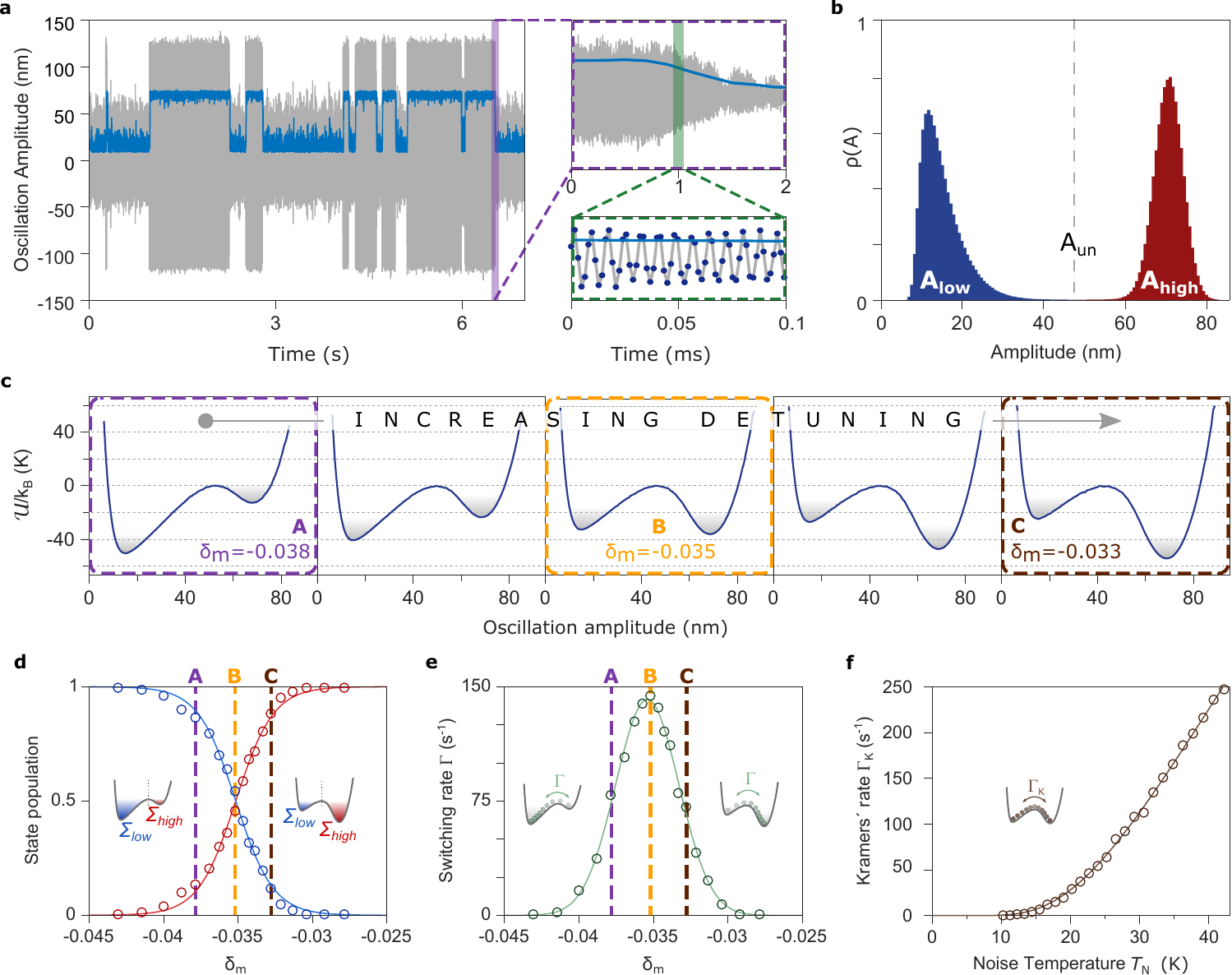}
\caption{\footnotesize \textbf{Stochastic switching and effective potential shaping.} (\textbf{a}), Position time trace (grey) and corresponding amplitude (blue) reveal a typical bistable behaviour of the system, with randomly distributed stochastic switches between the two stable oscillation states. The distinctive trait of our system is to be able to follow the dynamics of the overdamped variable down to very short time scales (insets) and therefore control the dynamics with high accuracy. (\textbf{b}) A histogram of the amplitude distribution featuring two fully separated amplitude states from which the effective potential $\mathcal{U}(A)$ can be retrieved by inverting the Boltzmann-Gibbs distribution. (\textbf{c}) With ad-hoc tuning of the driving parameters $(\delta_m,\epsilon)$ the potential can be finely shaped. For the dynamics here shown one expects a state population inversion and a maximization of the transition rate as shown in \textbf{d} and \textbf{e} respectively, where points A, B and C correspond to the marked potentials in \textbf{c}. (\textbf{f}) Transition rate as a function of noise injected for the symmetric potential case. Solid line is a fit to Kramers' law, which yields $\Gamma_0=(1814\pm96)~\mathrm{s}^{-1}$ and $\Delta \mathcal{U}/k_B = (83\pm2)~\mathrm{K}$.}
\label{fig:StochSwitch}
\end{figure*}
\newline \indent Inside the bistable region, natural thermal noise can activate spontaneous transitions between low and high amplitude states. However, in high vacuum these events are extremely rare, which makes our system a promising candidate for nonlinear sensing. In order to observe stochastic activation within a reasonable time we therefore add to the EOM a Gaussian noise $\zeta(t)$ of amplitude $N_V$, that in turns produces a position dependent parametric stochastic force of the form $\mathcal{F}\und{noise} \propto N_V \zeta(t) x$. Noise calibration can be performed by looking at the particle's oscillation amplitude $A$ in the absence of parametric driving~\cite{Mestres2014Realization}. For low noise levels, the particle's dynamics remains unaffected, being still dominated by the thermomechanical noise of the environment. However, for sufficiently high noise, the particle's amplitude starts to increase. Assuming an effective temperature $T_{\mathrm{eff}}=\frac{m\Omega_0^2 A^2}{k_B}$, the noise amplitudes $N_V$ are calibrated into meaningful temperature units $T_{\mathrm{N}}$ (See Supplementary Note~5). Interestingly, the measured trend differs from what reported in other systems~\cite{Venstra2013Stochastic}, where a quadratic dependence between $T_N$ and $N_V$ was found. This is to be ascribed to the parametric (i.e. multiplicative) nature of the noise injected in the present case, and prompts the interest of future studies.
\newline \indent The stochastic switching dynamics of the levitated nanoparticle is investigated via the injection of noise. Figure ~\ref{fig:StochSwitch}a illustrates position and amplitude time traces corresponding to few switching events and emphasizes the remarkable ability of our tracking scheme to resolve considerably different time scales. 
The amplitude probability distribution $\rho(A)$, obtained by analysing the particle trajectory over approximately one minute, is shown in Fig.~\ref{fig:StochSwitch}b. 
Even though the particle is in a non-thermal state due to the applied  parametric control~\cite{Gieseler2014Dynamic}, we introduce a simplified model that describes the dynamics of its amplitude as the motion of a thermal fictitious particle in a double well potential. This justifies the use of the Boltzmann-Gibbs distribution $\rho(A) \propto \mathrm{exp} \left[ - \frac{\mathcal{U}(A)}{k\und{B} T_N} \right]$ to extract the amplitude effective potential  $\mathcal{U(A)}$ that models the bistable dynamics. Note that this approximation does not induce any inconsistency in the subsequent analysis, which relies in the experimental determination of the symmetric condition for $\mathcal{U(A)}$.
The high frequency stability of the system allows us to finely modify the probability distribution $\rho(A)$, and hence to engineer almost at will the shape of $\mathcal{U}(A)$ following any path inside the bistable regime. An interesting case consists of the so called iso-amplitude lines: paths along which the amplitude $\mathcal{A}\und{high}$, and consequently the amplitude gap $\mathcal{A}_{\Delta} = \mathcal{A}_\mathrm{high}-\mathcal{A}_\mathrm{low}$, is kept constant. These particular subsets of the phase space are nonlinear features generally hidden by the frequency noise that blurs the measured instability tongue~\cite{Almog2007Signal,Gieseler2014Nonlinear}. Thanks to the reduced frequency fluctuations, they are clearly visible in our system. Few examples of iso-amplitude lines are shown in the lower panel of Fig.~\ref{fig:Exp_SetUp}d (solid lines), and satisfy the following relation (See Supplementary Note~4 for derivation):
\begin{equation} \label{eq:iso-amplitude}
\epsilon(\delta\und{m}) = \left[ \left( 1+\frac{9 \xi^2}{\eta^2} \right)\frac{\delta\und{m}^2}{\delta\und{th}^2}-6\xi \mathcal{A}_{\Delta}^2 \delta\und{m} + \eta^2 \delta\und{th}^2 \mathcal{A}_{\Delta}^4 \right]^{1/2}
\end{equation}
being $\delta\und{th}=\sqrt{9\xi^2+\eta^2}/2\eta$. By following an iso-amplitude line with a fixed $\mathcal{A}_{\Delta}$, we obtain a very smooth evolution of the effective potential (see Fig.~\ref{fig:StochSwitch}c). Starting from an asymmetrically tilted configuration, it progressively undergoes inversion of its shape, passing through a quasi-symmetric condition where the potential barrier determines equal depth of the two wells. It is important to stress that, upon this ad-hoc dynamical sweep of $\epsilon$ and $\delta_m$, the two minima of the potential corresponding to the amplitudes $\mathcal{A}\und{high}$ and $\mathcal{A}\und{low}$, maintain their position fixed. Clearly, this would not be the case when following any other path, for example moving along $\epsilon=\mathrm{const}$ lines (see cross-cuts in Fig.~\ref{fig:Exp_SetUp}d). The populations of the two states $\Sigma_{\rm high}$ and $\Sigma_{\rm low}$, defined as the normalized integrals of the amplitude distribution respectively above and below a threshold $\mathcal{A}_{\mathrm{un}}$, are shown in Fig.~\ref{fig:StochSwitch}d, and display an inversion consistent with the potential dynamics observed in Fig.~\ref{fig:StochSwitch}c. The switching rate $\Gamma$ also depends significantly on the detuning (see Fig.~\ref{fig:StochSwitch}e), with a maximum corresponding to the symmetric potential configuration. Under this symmetry condition the switching rate, as a function of noise temperature $T_{N}$, is expected to follow Kramers' law $\Gamma_K=\Gamma_0\exp(\frac{-\Delta \mathcal{U}}{k_B T_N})$. However, the injection of noise and its interplay with nonlinearities leads to a drift of the oscillation frequency, equivalent to a change in the detuning $\dm$, thereby distorting the symmetric potential (see Supplementary Note~6). If this change were not taken into account, a decay of the switching rate would be observed~\cite{Venstra2013Stochastic}. Conversely, if an adjusted detuning $\dm^{*}(T_N)$ is applied (See Supplementary Fig.~8), the symmetric potential is preserved and the prediction of Kramers' law is perfectly met, as we show in Fig.~\ref{fig:StochSwitch}f.
\begin{figure*}[ht!]
\centering
\includegraphics[width=0.8\textwidth]{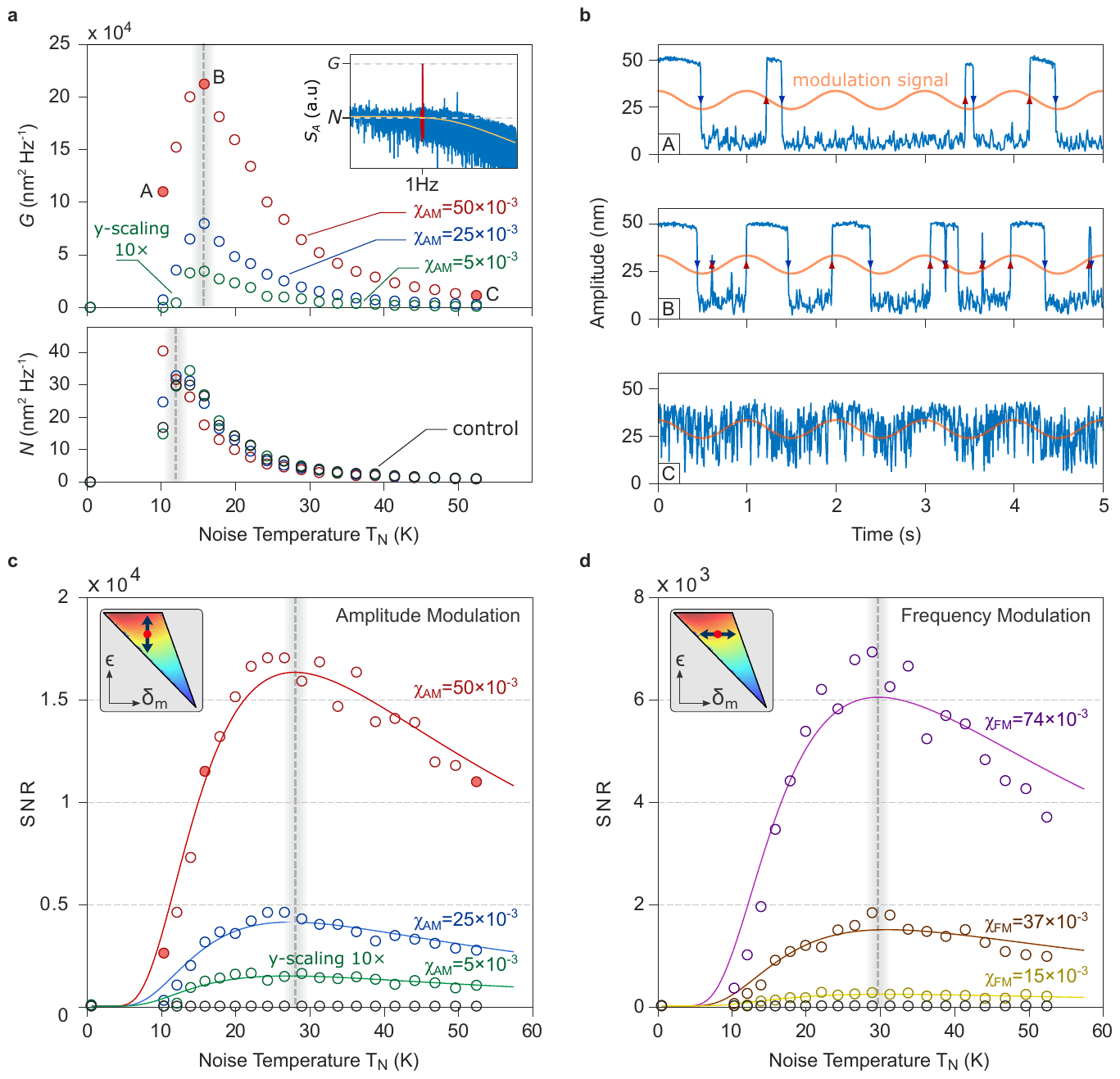}
\caption{\footnotesize \textbf{Stochastic resonance experiment} (\textbf{a}) Spectral amplification $G$ and noise floor $N$ as a function of the noise temperature for three different modulations. (\textbf{b}) Three examples of amplitude time traces for different noise temperatures (corresponding to points A,B and C in \textbf{a}), together with the modulation signal (orange line, not to scale in y axis). For low noise (A) there is little correlation between switching dynamics and modulating signal. However, for higher noise (B) the system reaches synchronization and a maximum in $G$ is found. If further increased (C), the noise leads to a degradation of the coherence in the switching dynamics. (\textbf{c}) The signal to noise ratio (SNR) of the detected modulating signal clearly presents a maximum (resonance) at $T_N\sim28~\mathrm{K}$. Note that the peak position depends on both $G$ and $N$ and is therefore not expected to coincide with the peak in \textbf{a}. (\textbf{d}) The SNR curve for the SR experiment performed with frequency modulation of the driving signal. The same observations reported for \textbf{c} apply here. In both cases, black circles represent control data with $\chi_{AM/FM} = 0$ to show that no amplification is encountered without modulation.}
\label{fig:StochRes}
\end{figure*}
\newline \indent The achieved control over the bistable dynamics makes of our system a very flexible platform for studying two different signal amplification schemes. The first relies on a common phenomenon in bistable systems called stochastic resonance (SR)~\cite{Gammaitoni1998Stochastic}. A weak periodic perturbation induces a modulation of the potential barrier separating the two states, and leads to an overall synchronized (i.e. quasi-coherent) switching dynamics when the interwell transition rate matches twice the frequency of the perturbation. Given that this rate depends monotonically on the amount of noise in the system, SR leads to a noise-induced rise (and then fall) of the signal-to-noise ratio (SNR), and can therefore be exploited to amplify a narrow-band signal in a nonlinear system under appropriate conditions.
\newline \indent To this aim, we prepare the bistable system driving the particle with suitably chosen parameters $\left(\delta_m^{*}, \epsilon^{*} \right)$. Once more, the noise-dependent detuning $\delta_m^{*}(T_N)$ avoids the escape from the bistable region as a consequence of hysteresis quenching~\cite{Aldridge2005Noise,Venstra2013Stochastic}, which in turn would prevent the observation of the full SR curve when noise is increased. The periodic ($\tilde{\omega}/2\pi = 1~\mathrm{Hz}$) perturbation is then introduced modulating the depth of the parametric driving signal, namely $\epsilon(t) = \epsilon^*\left[1+\chi_{\rm AM} \cos (\tilde{\omega} t )\right]$, where the modulation strength $\chi_{\rm AM}$ corresponds to optical forces of ten to a hundred atto-Newtons for the parameters used in our experiment. Amplitudes are measured with long acquisition times ($\sim 10^3~\rm{s}$) and repeated for increasing noise temperatures.
\newline \indent At $T_N \sim 10~\mathrm{K}$ (point marked A) switching events are only partially correlated with the modulation signal (see Fig.~\ref{fig:StochRes}b). Instead, when noise is increased to $T_N \sim 16K$ (point B) the system clearly exhibits an overall synchronization. A spectral analysis of the corresponding amplitude trace shows that $S_{A}$ features an extremely sharp peak precisely at $\tilde{\omega}/2\pi$ (see inset in Fig.~\ref{fig:StochRes}a), for which spectral amplification $G$ and noise floor $N$ can be defined. When noise is further increased, the switching dynamics loses correlation with the modulation signal. Interestingly, a similar and counterintuitive behaviour is observed in the noise floor $N$, which presents a similar non-monotonic trend caused by a redistribution of noise intensity towards higher frequencies. Linear perturbation theory of stochastic resonance~\cite{Gammaitoni1998Stochastic} (See Supplementary Note~7) predicts a $SNR=G/N$ of the form:
\begin{equation}\label{eq:SNR_theory}
SNR=\pi\left( \frac{\chi \mathcal{A}_{\Delta}}{T_N}\right)^2 \Gamma_0 \exp \left(-\frac{\Delta \mathcal{U}}{k_{B}T_N}\right) + \mathcal{O}(\chi^4).
\end{equation}
where $\chi$ is the modulation of the potential, proportional to $\chi_{AM}$.
\begin{figure}[ht!]
\centering
\includegraphics{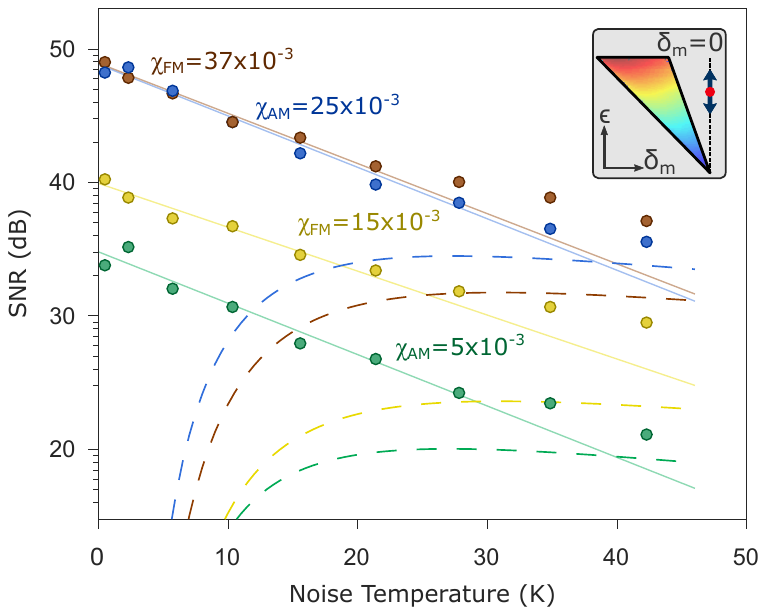}
\caption{\footnotesize \textbf{Direct amplification experiment.} SNR of a $1~\rm Hz$ weak modulation signal detected by the system prepared at resonance ($\dm=0$) outside of the bistable regime, but still inside the instability tongue (inset). Circles are experimental data points for different AM/FM runs. Solid lines are fitted exponential decays, while dashed lines correspond to the fitted functions~(\ref{eq:SNR_theory}) of Fig.~\ref{fig:StochRes}c,d and allow a straight comparison of the results obtained with the SR experiment (color coding preserved). In particular, independently of the noise level, direct amplification always features a higher SNR than the corresponding SR case, with the two curves approaching for high noise, in agreement with the central dogma of signal detection theory.}
\label{fig:DirectDet}
\end{figure}
Figure~\ref{fig:StochRes}c shows the experimental SNR curves for different values of $\chi_{\rm AM}$, together with a fitting of eq.~(\ref{eq:SNR_theory}) that displays very good agreement. Similarly, the ratios of the fitted modulation depths $(2:5.4:11)$ are also in good agreement with the expected ratios $(2:5:10)$.
\newline \indent A second SR experiment was carried out for frequency modulation of the driving signal, $\delta_m(t) = \delta_m^*\left[1+\chi_{\rm FM} \cos (\tilde{\omega} t )\right]$, which according to our study (see Fig.~\ref{fig:StochSwitch}c), induces a potential modulation consistent with SR requirements. The corresponding results are shown in Fig.~\ref{fig:StochRes}d. We emphasize the fact that previous experiments have only explored amplitude modulation, and that this is the first experimental demonstration of SR with frequency modulated signals. Again the fitted modulation strengths give ratios $(1.9:2.6:4.9)$, in good agreement with the expected ones $(2:2.5:5)$. Interestingly, the resonance appears at a noise temperature equivalent to the one observed in the amplitude modulation case. This is consistent with the fixed value of $\tilde{\omega}$ along the two experiments that requires an equal $\Gamma_K$ in order to fulfil the time-scale matching of the resonant condition.
\newline \indent Our platform enables us to directly compare the SR experiment with another amplification scheme that is implemented by changing the driving parameters to zero detuning. For $\delta_m = 0$, the particle is driven outside of the bistable regime, but still inside the instability tongue (see inset in Fig.~\ref{fig:DirectDet}). In this configuration, the effective potential is monostable, and the AM and FM modulations applied (having same $\chi_{AM/FM}$ as in the SR case) result in a modulation of the oscillation amplitude for which a spectral analysis still leads to the observation of a peak at $\tilde{\omega}$. The corresponding SNR as a function of noise temperature $T_N$ are shown in Fig.~\ref{fig:DirectDet}, together with fitted exponential decays that properly follow the experimental data for $T_N<30~\rm{K}$, and the fits of Fig.~\ref{fig:StochRes}c,d in order to ease a direct comparison of the two amplification schemes. Independently of the noise level, direct amplification always features a higher SNR than the corresponding SR, and the two methods give similar outcomes at high $T_N$. This result is an experimental verification of the central dogma of signal detection theory~\cite{Tougaard2002Signal}, namely that stochastic resonance can decrease the SNR degradation of a noisy signal but it does not provide a mechanism by which the undetectable becomes detectable~\cite{Petracchi2000What,Dykman1998What}. The couterinuitive nature of SR, in fact, mainly relies on suboptimal parameter ranges~\cite{McDonnell2009What}, here exemplified by the fact that the same modulation becomes remarkably detectable when the detuning is set to zero, without the need of adding noise. Nonetheless, the fact of SR being observed in a wide variety of settings poses the question of why this phenomenon is omnipresent and favoured by nature. We surmise that suboptimal balanced configurations are generally preferable when dealing with very complex systems, for which optimal conditions would be difficult to achieve. Thus, exploiting ambient noise and stochastic resonance appears to be a successful strategy to ensure a robust amplification method that, at least above certain thresholds, is less sensitive to noise changes than other detection schemes (check Fig.~\ref{fig:DirectDet}).

\vspace{.5cm}
\noindent{\bf Discussion}
\newline \indent In conclusion, we presented an extensive study of the stochastic bistable dynamics of a levitated nanoparticle in high vacuum. The high stability achieved - in particular close to the bifurcation point - enabled the application of our system as a test platform for different amplification schemes, leading (to our knowledge) to the first qualitative and quantitative agreement between the stochastic resonance of a nanomechanical resonator and the corresponding analytical models. The present state of the experiment allowed to up to $\sim50~\rm dB$ amplification of a parametric force at the attonewton scale that was mimicked by modulating the trap stiffness. A similar perturbation could arise from a time varying nonlinear potential. However, to evaluate the performance of our system to detect a conventional linear force, a direct actuation is required. This could be implemented by electrically coupling a charged particle~\cite{Alda2016Trapping} to an external field~\cite{Ranjit2016Zeptonewton} or via scattering force from a weakly focused beam and is the subject of future work. The unprecedented performances demonstrated could enable the realization of novel ultra-sensitive threshold sensors, capable of detecting tiny perturbations via a state change in the system, or other detection schemes based on nonlinear nanomechanical resonators~\cite{Aldana2014Detection,Papariello2016Ultrasensitive}. Likewise, a high-Q parametrically driven Duffing resonator could boost the state-of-the-art in nanomechanical memory elements~\cite{Mahboob2014Two-mode,Bagheri2011Dynamic} introducing additional bits by simultaneous manipulation of the orthogonal oscillation modes~\cite{Frimmer2016Cooling}. In the parametrically driven regime our system could also enable fundamental search for classical to quantum transitions~\cite{Katz2007Signatures}, and prompt the realization of quantum-enhanced sensing techniques. Finally, we foresee in optically levitated nanoparticles a suitable platform for mimicking very complex stochastic nonlinear dynamics, able to shine a light on natural phenomena such as bio-molecule folding~\cite{Hayashi2012Single,Angeli2004Detection}, hearing~\cite{Eguiluz2000Essential,Martignoli2013Pitch} and neural signalling~\cite{McDonnell2011TheBenefits}.
\newline
\newline
\newline
\noindent\textbf{Methods}
\newline
\newline
\textbf{\small Sampling}. The amplitude $A(t)=\sqrt{\langle x^2 \rangle _\mathrm{n}}$ is tracked by sampling position $x$ at $f_s=625~\mathrm{kHz}$ and integrating over $\mathrm{n}$ successive positions measurements. The value of $\mathrm{n}$ depends on the temporal resolution needed in the particular measurement to be carried out. We set $\mathrm{n}=8192$ for the data shown in Fig.~\ref{fig:Exp_SetUp}, $\mathrm{n}=128$ for the amplitude time traces (blue data) of Fig.~\ref{fig:StochSwitch}a and to $\mathrm{n}=4096$ for the SR experiment of Fig.~\ref{fig:StochRes}. In this latter data sets, the acquisition time of a single amplitude trace at constant noise temperature (see Fig.~\ref{fig:StochRes}b) was $10^3~\rm s$. However, in order to display a meaningful dynamics, we show only $5~\rm s$ time traces.
\newline
\newline
\noindent\textbf{Data availability}
The data that support the findings of this study are available from the corresponding author on reasonable request.
\newline
\newline
\noindent\textbf{Authors' Present address}
\newline \noindent 
R.A.R.: \textit{Departamento de F\'isica At\'omica, Molecular y Nuclear, Universidad de Granada, 18071, Granada.}
\newline
M.S.: \textit{Center for Solid State Physics and New Materials, Institute of Physics in Belgrade, 11080 Belgrade, Serbia.}
\newline
L.R.: \textit{Laboratoire Aimé Cotton, CNRS, Université Paris-Sud, ENS Cachan,
Université Paris-Saclay, 91405 Orsay Cedex, France} 
\newline
\newline
\noindent\textbf{Corresponding authors}
\newline
F.R.: francesco.ricci@icfo.eu
\newline
R.Q.: romain.quidant@icfo.eu
%
%
%

%
%
%
\noindent\textbf{Acknowledgments}
\newline
The authors acknowledge financial support from the ERC- QnanoMECA (Grant No. 64790), the Spanish Ministry of Economy and Competitiveness, under grant FIS2016-80293-R and through the ``Severo Ochoa'' Programme for Centres of Excellence in R\&D (SEV-2015-0522), Fundació Privada CELLEX and from the CERCA Programme/Generalitat de Catalunya.  J. G. has been supported by H2020-MSCA-IF-2014 under REA grant Agreement No. 655369. L.R acknowledges support from an ETH Marie Curie Cofund Fellowship.
\newline
\newline
\noindent\textbf{Author contributions}
F.R., M.S., R.A.R. and R.Q. conceived the experiment. J.G. and F.R. designed and implemented the experimental set-up and wrote all data acquisition software. F.R. performed the experiment. F.R. and R.A.R. analysed the data with input from M.S., J.G., L.R. and L.N. All authors contributed to manuscript writing. R.Q. and L.N. supervised the work.

\end{document}